\def\year{2022}\relax
\documentclass[letterpaper]{article} 
\usepackage{aaai22}  
\usepackage{times}  
\usepackage{helvet} 
\usepackage{courier}  
\usepackage[hyphens]{url}  
\usepackage{graphicx} 
\usepackage{natbib}  
\usepackage{caption} 
\usepackage{amsmath}
\usepackage{booktabs}
\usepackage{array}
\newcolumntype{L}[1]{>{\raggedright\let\newline\\\arraybackslash\hspace{0pt}}m{#1}}
\newcolumntype{C}[1]{>{\centering\let\newline\\\arraybackslash\hspace{0pt}}m{#1}}
\newcolumntype{R}[1]{>{\raggedleft\let\newline\\\arraybackslash\hspace{0pt}}m{#1}}

\usepackage{xcolor}

\title{Limits of Multilayer Diffusion Network Inference in Social Media Research}
\author{
    Yan Xia,\textsuperscript{\rm 1}
    Ted Hsuan Yun Chen,\textsuperscript{\rm 1,2}
    Mikko Kivel\"{a}\textsuperscript{\rm 1} \\
}
\affiliations{
    \textsuperscript{\rm 1}Department of Computer Science, Aalto University, Espoo, Finland \\
    \textsuperscript{\rm 2}Faculty of Social Sciences, University of Helsinki, Helsinki, Finland \\
    yan.xia@aalto.fi, ted.hsuanyun.chen@gmail.com, mikko.kivela@aalto.fi

}

\begin{document}

\maketitle

\begin{abstract}
Information on social media spreads through an underlying diffusion network that connects people of common interests and opinions. This diffusion network often comprises multiple layers, each capturing the spreading dynamics of a certain type of information characterized by, for example, topic, language, or attitude. Researchers have previously proposed methods to infer these underlying multilayer diffusion networks from observed spreading patterns, but little is known about how well these methods perform across the range of realistic spreading data. In this paper, we conduct an extensive series of synthetic data experiments to systematically analyze the performance of the multilayer diffusion network inference framework, under varied network structure (e.g. density, number of layers) and information diffusion settings (e.g. cascade size, layer mixing) that are designed to mimic real-world spreading on social media. Our results show extreme performance variation of the inference framework: notably, it achieves much higher accuracy when inferring a denser diffusion network, while it fails to decompose the diffusion network correctly when most cascades in the data reach a limited audience. In demonstrating the conditions under which the inference accuracy is extremely low, our paper highlights the need to carefully evaluate the applicability of the inference before running it on real data. Practically, our results serve as a reference for this evaluation, and our publicly available implementation, which outperforms previous implementations in accuracy, supports further testing under personalized settings.
\end{abstract}

\section{Introduction}
Social media is a major channel of information diffusion \cite{guille2013information}, and thus provides rich data for understanding the crowd. On item-sharing-based platforms such as Twitter, item spreading patterns can reveal nuanced dynamics of both human-human interaction \cite{starbird2012will} and human-information interaction \cite{friggeri2014rumor, vosoughi2018spread}. While the who-to-whom spreading traces of items are usually not directly observable, researchers have proposed methods for inferring the underlying diffusion network among users from item spreading logs \cite{gomez2012inferring, myers2010convexity, gomez2011uncovering, gomez2013structure, du2012learning, rong2016model}. 


Earlier studies share the assumption of a single layer diffusion network. However, item spreading dynamics among the same set of actors can vary significantly depending on the topic, language, attitude, or other features of the item. This heterogeneity can be best represented with a multilayer network \cite{kivela2014multilayer} where node existence, edge connectivity, and edge weights can differ across layers. A multilayer diffusion network inferred from social media spreading data can potentially reveal exceptionally interesting dynamics, especially when it groups together cascades that spread similarly (i.e., on the same layer) and share more nuanced characteristics than topic or language.

While researchers have developed methods for inferring multilayer diffusion networks from spreading data \cite{wang2014mmrate, yang2014cluster, he2015hawkestopic, liao2016uncovering, he2017not, suny2018inferring}, these methods have only been tested under limited synthetic settings that likely differ from real-world situations. It is therefore near impossible to conduct empirical analyses using these methods, given the difficulty of assessing the accuracy of their results when applied to real data with unknown ground truth.

In this study, we bridge this gap by systematically testing how well the multilayer diffusion network inference framework performs under varied network and diffusion settings designed to cover a wide range of realistic social media spreading conditions. 
For each controlled feature, including network size, network density, number of layers, layer overlap, cascade size, and layer mixing, we report the extent to which the framework's performance varies.
Our results show that the accuracy of the multilayer decomposition is heavily dependent on multiple factors, including the density of the diffusion network and the size distribution of cascades in the spreading data. Notably, the inference framework fails to infer an accurate multilayer diffusion network when the underlying diffusion network is sparse or when most cascades in the data do not reach a large audience.



Our work delineates the limits of the multilayer diffusion network inference framework in potential social media applications. It highlights the need for researchers to evaluate the feasibility of the inference framework before applying it to real data, to avoid interpreting incorrect results and reaching invalid conclusions. The results of our experiments serve as a reference for this applicability evaluation by providing an estimate of the inference accuracy given the properties of the dataset. Although our results cannot cover all possible network structure and information diffusion settings, our open-source implementation\footnote{We have published our code at https://github.com/ecanet-research/multic.}, which achieves higher inference accuracy than previous implementations in comparable runtime, can be easily adapted for further testing under personalized setups.

\section{Related Work}
In the area of diffusion network inference, earlier studies mainly focus on the task of inferring a single layer network of information diffusion when the activation times of individuals are available but the spreading traces are not. Among them, \textsc{NetInf} \cite{gomez2012inferring} is one of the pioneering frameworks. In this study, the authors formulate the problem of estimating the diffusion network that maximizes the cascade likelihood under the Independent Cascade Model \cite{kempe2003maximizing} of cascade transmission, and offer an approximated solution based on submodular optimization. \textsc{ConNIe} \cite{myers2010convexity} deals with a similar inference problem but assumes a different prior for the edge transmission probability of each edge, and uses convex programming to optimize the objective, with an explicit $l1$-penalty term that induces sparsity. \textsc{NetRate} \cite{gomez2011uncovering} further assumes a different edge transmission rate for each edge under a continuous transmission time model, and defines a convex objective function with an inherent $l1$-penalty so that no manual hyperparameter tuning is needed to select the appropriate level of penalty. Other works further extend the \textsc{NetRate} framework, including \textsc{InfoPath} \cite{gomez2013structure} that infers a dynamic diffusion network which changes over time, and \textsc{KernelCascade} \cite{du2012learning} that uses kernel methods to support heterogeneous transmission time distributions beyond the assumed exponential, power law, or Rayleigh form in \textsc{NetRate}. Different from all approaches above, \citeauthor{rong2016model} (\citeyear{rong2016model}) proposed a completely model-free method that infers a diffusion network by clustering the cumulative distribution functions of transmission time intervals.

On top of the single layer inference methods, researchers have developed methods for inferring multilayer diffusion networks. \citeauthor{du2013uncover} (\citeyear{du2013uncover}) proposed \textsc{TopicCascade} that infers diffusion networks with topic-dependent transmission rates, but their model infers a cascade's topic distribution from its content. 
\citeauthor{wang2014mmrate} (\citeyear{wang2014mmrate}, \textsc{MMRate}) and \citeauthor{yang2014cluster} (\citeyear{yang2014cluster}, \textsc{MixCascades}) were among the first to build a general framework that infers multilayer diffusion networks solely from spreading data. 
Later, Liao, Chou, and Chen (\citeyear{liao2016uncovering}) proposed \textsc{FASTEN} that improves the inference accuracy by incorporating a decay parameter in the diffusion model. Another relevant line of research uses the Marked Multivariate Hawkes Process \cite{liniger2009multivariate} to infer multilayer diffusion networks. This includes \textsc{HawkesTopic} \cite{he2015hawkestopic}, \textsc{MultiCascades} \cite{he2017not}, and \textsc{MDM} \cite{suny2018inferring}. 

These studies on multilayer diffusion network inference mostly demonstrate the accuracy of the methods with synthetic data experiments. However, the range of synthetic test settings, as reported in the papers, is often limited. For example, the tests are usually performed on synthetic networks with a fixed number of nodes, edges, and layers \cite{wang2014mmrate, yang2014cluster, liao2016uncovering}. These settings also likely differ from real-world situations in, for example, having independently generated network layers and the assumption that any item spreads on either one layer or another. Since these network and diffusion settings can significantly affect the inference accuracy, the applicability of the inference framework is largely unknown under different, more realistic circumstances. We contribute to this knowledge in our work through a systematic testing of the framework.

\section{Testing Setup}
To systematically assess the limits of the multilayer diffusion network inference framework, we tested its performance under varied network and diffusion settings using a series of synthetic data experiments. In each experiment, we varied a single parameter of the synthetic data generation process and observed how the accuracy of the inference framework on synthetic data changes with the value of the parameter. To cover a relatively wide range of realistic spreading conditions, we varied cascade size distribution, cascade filtering, network density, network size, number of layers, layer overlap, and mixed spreading. In this section, we outline our testing setup.

\subsection{Inference Framework}
We start by building an effective and efficient implementation of the inference framework to be tested. To the best of our knowledge, \textsc{MMRate} \cite{wang2014mmrate} is the most recognized framework designed for multilayer diffusion network inference from spreading data. Therefore, we use for our testing an inference framework that has a similar problem formulation and inference method as in \textsc{MMRate}, but apply a more effective optimization method in our implementation. We will demonstrate in \textit{Validation Test} the performance of our implementation over other public implementations of multilayer diffusion network inference, including \textsc{MMRate}, \textsc{MixCascades} \cite{yang2014cluster} and \textsc{FASTEN}\footnote{All these implementations are provided at \url{https://github.com/plliao/FASTEN}.} \cite{liao2016uncovering}.

\subsubsection{Problem Formulation}
We formulate the problem of multilayer diffusion network inference as follows. Suppose there exists among a set of $N$ nodes (i.e., users) an underlying directed multilayer diffusion network of $K$ layers, $G=\{(V^k, E^k)|k=1\dots K\}$, where $V^k$ is the set of nodes on layer $k$, and $E^k$ is the set of edges on layer $k$. We define $\alpha_{ij}^k\in[0,1]$ as the edge weight (or edge transmission rate) from node $i$ to node $j$ on layer $k$. Here, we assume $\alpha_{ij}^k$ is defined for all $i$, $j$, and $k$, but a directed edge $(i,j)$ exists in $E^k$ if and only if $\alpha_{ij}^k\neq 0$, and a node $i$ exists in $V^k$ if and only if $i$ is not an isolated node on layer $k$. Note here that $\boldsymbol{\alpha}$ stores complete information on edge connectivity, node existence, and the exact edge weights, so the inference of $\boldsymbol{\alpha}$ implicitly includes the inference of $G$. 

We then define a cascade $c$ as an item (e.g., tweet, post, hashtag, news article) that spreads on this network. We denote $\pi_k^c\in [0,1]$ as the probability that cascade $c$ spreads on layer $k$ (later referred to as the ``layer membership parameter''), with $\sum_1^K \pi_k^c=1$ for all $c$. In real-world spreading data, we can neither directly observe the edge transmission rates $\boldsymbol{\alpha}$ nor the cascade layers $\boldsymbol{\pi}$. Instead, we observe for each cascade $c$ a set of user activation logs $\{t_1^c, \dots, t_N^c\}$: here we let $t_n^c$ be the time node (i.e., user) $n$ gets activated on cascade $c$ (e.g., the time $n$ retweets the tweet) if they are activated, or the ending time $T$ if they are not activated before $T$. The task is to infer $\boldsymbol{\alpha}$ and $\boldsymbol{\pi}$ from the user activation logs we observe.

\subsubsection{Inference Method}
We conduct the inference by assuming a generative diffusion model parameterized by $\boldsymbol{\alpha}$ and $\boldsymbol{\pi}$, and finding the values of $\boldsymbol{\alpha}$ and $\boldsymbol{\pi}$ that maximize the likelihood of the observed spreading data under the assumed diffusion model. More specifically, we adopt a continuous transmission time diffusion model and a survival analysis framework for computing the likelihood of data, as first proposed in \textsc{NetRate} \cite{gomez2012inferring} and generalized in \textsc{MMRate}.

Formally, let $\Delta t_{ij}^c=t_j^c-t_i^c$ denote the transmission time of cascade $c$ from node $i$ to node $j$, namely the difference between the activation times of $i$ and $j$ on $c$. We assume $\Delta t_{ij}^c$ follows the exponential distribution\footnote{The framework potentially generalizes to other distributions, yet for simplicity we will assume exponential form throughout this work.} parameterized by $\lambda_{ij}^c$, where $\lambda_{ij}^c=\sum_{k=1}^K \pi_k^c \alpha_{ij}^k$ is the sum of edge transmission rates across all layers, weighted by the layer membership parameters $\pi_k^c$ of cascade $c$. In other words, we assume the probability of cascade $c$ successfully spreading from node $i$ to node $j$ with time interval $\Delta t_{ij}^c$ to be $f(\Delta t_{ij}^c;\lambda_{ij}^c)$, where $f(\cdot)$ is the probability density function (PDF) of the exponential distribution parameterized by $\lambda_{ij}^c$. Intuitively, the probability of observing a shorter transmission time $\Delta t_{ij}^c$ increases with $\lambda_{ij}^c$.

We then define the failure probability of transmissions using the survival function of the transmission time distribution, $S(t;\lambda)=\int_t^{\infty} f(x;\lambda)dx$. Given a node $i$ that is activated on cascade $c$ at time $t_i^c$, and a cascade-specific edge transmission rate $\lambda_{ij}^c$ from node $i$ to node $j$, we assume the probability that $j$ is not activated on $c$ until time $T$ to be $S(T-t_i^c;\lambda_{ij}^c)$. Intuitively, this is the probability that $j$ survives the activation from $i$ on $c$ until time $T$.

From this, we get the likelihood of observing the cascade spreading logs under this generative model. Specifically, the likelihood of observing node $j$ activated on cascade $c$ by a previous node $i$ will be 
\small
$$\Gamma_{ij}^{+}(c)=f(\Delta t_{ij}^c;\lambda_{ij}^c) \prod_{u:u\neq i,t_u^c<t_j^c} S(\Delta t_{uj}^c;\lambda_{uj}^c)\,,$$
\normalsize
which is the probability that node $j$ is activated by exactly node $i$ and survives activations from all other nodes that are activated earlier than node $j$. Then, the likelihood of observing node $j$ activated on cascade $c$ by any previous node is the sum of the above likelihood over all possible $i$'s:
\small
$$\begin{aligned}
\Gamma_j^+(c)&=\sum_{i:t_i^c<t_j^c} \Gamma_{ij}^{+}(c)\,.
\end{aligned}$$
\normalsize
On the other hand, the likelihood of a node $n$ not being activated on cascade $c$ by the ending time $T$ is
\small
$$\Gamma_n^-(c)=\prod_{m:t_m^c<T} S(T-t_m^c;\lambda_{mn}^c)\,,$$ 
\normalsize
which is the probability that it survives all possible activations.
The likelihood of observing the entire activation sequence of $c$, is then the joint likelihood of observing all the successful activations and all the failed activations\footnote{We include all derivations in \textit{Derivation of Formulas}.}:
\small
$$\begin{aligned}
L(c)=&\prod_{j:t_j^c<T} \Gamma_j^+(c) \times \prod_{n:t_n^c>T} \Gamma_n^-(c) \\
=&\prod_{j:t_j^c<T} \left[\prod_{u:t_u^c<t_j^c} S(\Delta t_{uj}^c;\lambda_{uj}^c)\times \sum_{i:t_i^c<t_j^c} H(\Delta t_{ij}^c;\lambda_{ij}^c) \right. \\
&\times \left. \prod_{n: t_n^c>T} S(T-t_j^c;\lambda_{jn}^c)\right],
\end{aligned}$$
\normalsize
where $H(t;\lambda)=f(t;\lambda)/S(t;\lambda)$ is the hazard function of the transmission time distribution.

The total likelihood of all cascades is then $\prod_c L(c)$. When $f(\cdot)$ is the PDF of the exponential distribution, we can write the negative log likelihood of all cascades as 
\small
$$\begin{aligned}
\mathcal{L}(\mathbf{c};&\boldsymbol{\alpha}, \boldsymbol{\pi})=\sum_c ( -\log L(c) )\\
=&\sum_c \sum_{j:t_j^c<T}\left(\sum_{u:t_u^c<t_j^c} \sum_{k=1}^K \Delta t_{uj}^c\pi_k^c \alpha_{uj}^k \right.\\
&-\left. \log\sum_{i:t_i^c<t_j^c} \sum_{k=1}^K \pi_k^c \alpha_{ij}^k + \sum_{n: t_n^c>T}\sum_{k=1}^K (T-t_j^c) \pi_k^c \alpha_{jn}^k \right).
\end{aligned}$$
\normalsize

The goal of the inference is then to find the values of $\boldsymbol{\alpha}$ and $\boldsymbol{\pi}$ within the constraints that maximize the total likelihood of all cascades, or equivalently, minimize the total negative log likelihood $\mathcal{L}(\mathbf{c};\boldsymbol{\alpha}, \boldsymbol{\pi})$. More formally, the inference problem corresponds to the constrained optimization problem
\small
$$\begin{aligned}
\underset{\boldsymbol{\alpha}, \boldsymbol{\pi}}{\text{minimize}}\quad &\mathcal{L}(\mathbf{c};\boldsymbol{\alpha}, \boldsymbol{\pi}) \\
\text{subject to\quad} & 0\leq \pi_k^c\leq 1 \text{ for all } k \text{, for all } c, \\
& \sum_1^K \pi_k^c=1 \text{ for all }c, \\
& 0\leq \alpha_{ij}^k\leq 1 \text{ for all } i \text{, for all } j \text{, for all } k.
\end{aligned}$$
\normalsize

\subsubsection{Implementation}
In our model, we have continuous constraints of $\boldsymbol{\pi}$ as opposed to discrete ones in the \textsc{MMRate} model (i.e., $\pi_k^c\in \{0,1\}$). This offers us a wider choice of modern optimization tools that can potentially solve the problem more efficiently. In specific, we use the PyTorch Python library \cite{paszke2019pytorch} because it is known for supporting fast computation and automatic differentiation of heavy optimization problems using graphics processing units (GPUs). 

To make our optimization problem efficiently solvable with PyTorch routines, we first transform the problem into an unconstrained one by replacing the variable constraints with variable transformations. Specifically, we first define unconstrained variables $\hat{\alpha}_{ij}^k$ for all $i,j,$ and $k$, and replace every $\alpha_{ij}^k$ with $\sigma(\hat{\alpha}_{ij}^k)$ in the objective function, where $\sigma(x)=1/(1+e^{-x})$ is the sigmoid function that converts any input from $(-\infty, \infty)$ to the $(0,1)$ interval. Similarly, we define unconstrained variables $\hat{\pi}_k^c$ for all $c$ and all $k\in\{1,2,\dots,K-1\}$. To satisfy the constraints of $\sum_1^K \pi_k^c=1$, we let 
\small
$$\begin{aligned}
\pi_1^c =& \sigma(\hat{\pi}_1^c), \\ 
\pi_2^c =& \sigma(\hat{\pi}_2^c)(1-\pi_1^c), \\
&\vdots \\
\pi_{K-1}^c =& \sigma(\hat{\pi}_{K-1}^c)(1-\sum_{k=1}^{K-2} \pi_k^c), \\
\pi_K^c =& 1-\sum_{k=1}^{K-1} \pi_k^c.\\
\end{aligned}$$
\normalsize
In this way, all constrained variables in the objective function can be converted from unconstrained ones, and the function can then be efficiently optimized using PyTorch.

To take full advantage of GPU resources that accelerate large matrix computations significantly, we rewrite the objective function in matrix form. Suppose $C$ is the number of cascades in the dataset. Let $\Pi$ be a $C\times K$ matrix where $\Pi_{ck}=\pi_k^c$, $A$ be a $K\times N\times N$ matrix where $A_{kij}=\alpha_{ij}^k$, and $\Delta T$ be a $C\times N\times N$ matrix where 
\small
$$
\Delta T_{cij}=
\begin{cases}
    \Delta t_{ij}^c,& \text{if } t_i^c<t_j^c<T\\
    T-t_i^c,& \text{if } t_i^c<T, t_j^c=T\\
    0,              & \text{otherwise}
\end{cases}
$$
\normalsize
Additionally, let $M$ be a $C\times N\times N$ mask matrix where
\small
$$
M_{cij}=
\begin{cases}
    1,& \text{if } t_i^c<t_j^c<T\\
    0,              & \text{otherwise}
\end{cases}
$$
\normalsize
Then the objective function can be written as
\small
$$\text{sum}(\Delta T\odot (\Pi A))-\text{sum}(\text{log}(M\odot (\Pi A))),$$
\normalsize
where $\odot$ denotes the element-wise matrix multiplication, $\text{sum}(X)$ denotes the sum of all elements in matrix $X$, and $\text{log}(X)$ denotes the element-wise $\log$ operation of matrix $X$ where zero elements are preserved. 

While this considerably speeds up the execution time, the memory consumption turns out to be a bottleneck: 
the program requires $O(N^2\cdot C)$ memory space on GPU, which makes it difficult to scale to large networks. 
To mitigate the memory consumption issue, we split the inference into two phases. In the single layer phase, we infer which edges exist on any layer of $G$. Or, in other words, we infer which edges exist in the aggregated single layer network $G_A=(V_A, E_A)$, where $V_A=\bigcup_{k=1}^K V^k$ is the aggregation of all nodes in all layers of $G$, and $E_A=\bigcup_{k=1}^K E^k$ is the aggregation of all edges in all layers of $G$. The set of edges inferred to exist in the aggregated network is denoted as $E_S$. Then, in the multilayer phase, we infer the layer-wise edge transmission rates $\boldsymbol{\alpha}$ within $E_S$ and cascade layers $\boldsymbol{\pi}$.

Moreover, we notice that it is unlikely for an edge to exist between two nodes that never occur in the same cascade. Therefore, in the single layer phase, we only consider the set of ``possible'' edges -- i.e., edges between nodes that co-occur in at least one cascade -- and denote this set as $E_P$. We include a formulation of the two inference phases in \textit{Formulation of Inference Phases}.

In the improved implementation, the single layer inference uses $O(\max(|E_P|, N\cdot C))$ memory, and the space complexity of the multilayer inference reduces from $O(N^2\cdot C)$ to $O(|E_S|\cdot C)$.

\subsection{Data}
\begin{figure}[t]
    \centering
    \includegraphics[width=\linewidth]{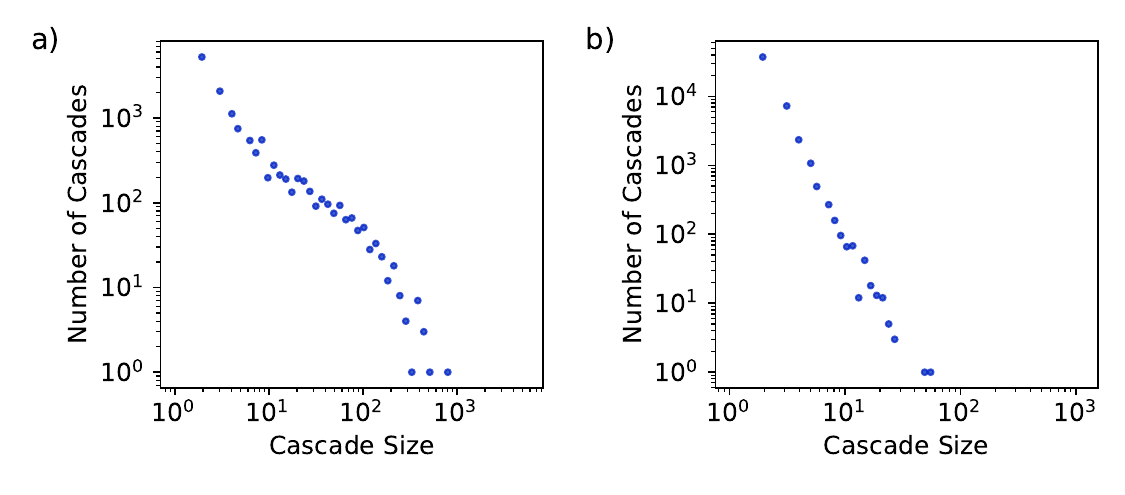}
    \caption{Cascade size distributions of the two real-world datasets in log-log scale. Plot a) corresponds to the \textit{ClimateSkepticCascades} dataset, and plot b) corresponds to the \textit{PoliSciCascades} dataset.}
    \label{fig:rt}
\end{figure}

To conduct a systematic analysis of the inference framework, we need to generate synthetic spreading data under different realistic settings. To inform this data generation, we collected real-world datasets of information diffusion on social media. We first built the \textit{ClimateSkepticCascades} dataset from the Twitter climate discussion dataset \cite{xia2021spread}. The original dataset contains all climate-related tweet, retweet, or reply records during the announcement of the 2019 Nobel Peace Prize, where the authors recognize a division of climate activists and climate skeptics in the retweet network. Observing interesting discussion dynamics among the climate skeptics, we filtered the original dataset to a subset with only nodes and records from the skeptic group. The resulting \textit{ClimateSkepticCascades} dataset contains 5816 nodes (i.e., users) and 41385 cascades, with each cascade corresponding to the spreading trace of one original tweet. Among them, 13007 have at least one retweet. Figure~\ref{fig:rt}a shows the cascade size distribution in log-log scale. The distribution indicates that the dataset contains a fair number of cascades with moderate size, but the total number of cascades is rather small compared to the number of nodes. 

We then collected a second dataset where nodes in the underlying network belong to a community that is relatively stable and well-connected, instead of purely topic-induced. Specifically, we built the \textit{PoliSciCascades} dataset through the following process: first, we extracted the Twitter handles of 1236 political science professors at PhD-granting institutions in the United States from the \#polisci Twitter dataset \cite{bisbee2020polisci}; then, we used Twitter's timeline v1.1 API endpoint\footnote{\url{https://developer.twitter.com/en/docs/twitter-api/v1/tweets/timelines/api-reference/get-statuses-user_timeline}} to fetch the most recent timeline records of each user, including up to 3200 tweet, retweet, or reply records per user before April 8, 2021. The resulting \textit{PoliSciCascades} dataset contains 1158 nodes and 1618544 cascades, of which 49255 have at least one retweet. Figure~\ref{fig:rt}b shows the log-log scale cascade size distribution. Overall, there is a large number of cascades circulating within the community, but most cascades have a small number of retweets with respect to the total number of users in the dataset. 

\subsubsection{Generating the Network}
Given the reference of real spreading data, we design a set of controlled settings for generating synthetic data. To begin with, we use the directed configuration model \cite{newman2001random} -- that produces random directed networks with designated in-degree and out-degree sequences -- to generate the ground-truth diffusion networks that we need to infer; notably, such degree-driven approach for generating synthetic networks is widely used in previous studies of spreading processes on multilayer networks \cite{salehi2015spreading}. We randomly generate the in-degree and out-degree sequences that feed into the configuration model by sampling from log-normal distributions, based on the fact that log-normal models provide good fit to the degree sequences of the Tumblr reblog network \cite{xu2014rolling}. 

Considering the computational limits, we have the core network setting of $N=1000, K=2, \phi=0, \mu_{in}=0.5, \sigma_{in}=1, \mu_{out}=0, \sigma_{out}=\sqrt{2}$, where $N$ is the number of nodes in the network, $K$ the number of layers in the network, $\phi$ the edge overlap parameter between the layers, and $\mu_{in}$ (resp. $\mu_{out}$) and $\sigma_{in}$ (resp. $\sigma_{out}$) are the mean and standard deviation parameters of the log-normal distribution that we use to generate the in-degree (resp. out-degree) sequence for each layer of the network. For a multilayer network with $\phi=0$, we independently generate each layer using the directed configuration model. For each directed edge $(i,j)$ that exists on layer $k$ in the network, we generate its edge transmission rate $\alpha_{ij}^k$ by sampling uniformly from $(0.01, 1)$.

Beyond the core setting, we also generate networks of varied density, size, number of layers, and layer overlap, specifically under the four sets of settings below (parameters that have the same values as in the core setting are omitted):

\begin{enumerate}
    \item Varied network density:
    \begin{enumerate}
        \item $\mu_{in}=0, \sigma_{in}=1, \mu_{out}=0, \sigma_{out}=1$
        \item $\mu_{in}=1, \sigma_{in}=1, \mu_{out}=0, \sigma_{out}=\sqrt{3}$
    \end{enumerate}
    \item Varied network size: $N=2000, N=4000$
    \item Varied number of layers: $K=3, K=4, K=5$
    \item Varied layer overlap: $\phi=1, \phi=0.5$
\end{enumerate}

Note that under the setting of $\phi=1$, we first generate the first layer of the network and then simply copy the edge structure to the second layer, while the edge transmission rates are still sampled independently on each layer. Under the setting of $\phi=0.5$, we generate the first layer, copy the edge structure to the second layer, and then randomly rewire 50\% of the edges on the second layer. We allow self-loops and parallel edges when rewiring\footnote{If not, the rewiring process will be biased in that edges adjacent to nodes of higher degrees will have a smaller probability of being rewired successfully.} and remove them after the entire process, therefore the true edge overlap rate is slightly higher than 0.5; the actual overlap rate we got is around 0.58.

\subsubsection{Generating the Spreading Logs}
Given a synthetic diffusion network, we use the Gillespie algorithm \cite{gillespie1977exact} to generate the information cascades by simulating susceptible-infectious-removed (SIR) processes \cite{newman2002spread} on the network, under the SIR compartmental model that is extensively used in previous studies for simulating spreading processes on multilayer networks \cite{salehi2015spreading}.

Specifically, for each cascade $c$, we first sample $k_c$, the main layer it spreads on, uniformly from $\{1,2,\dots,K\}$. To allow a certain level of mixed spreading on multiple layers, so that the spreading of a cascade depends not only on the edge transmission rates on a single layer but a weighted sum of the rates on multiple layers, we additionally define a noise parameter $\epsilon_c$, such that $\pi_{k_c}^c=1-\epsilon_c$, and $\pi_k^c=\epsilon_c/(K-1)$ for all $k\neq k_c$. We sample $\epsilon_c$ for each cascade uniformly from $(0, \epsilon_{max})$, where we let $\epsilon_{max}$ be respectively 0, 0.2, or 0.4, to cover different levels of mixed spreading. 

\begin{figure}[t]
    \centering
    \includegraphics[width=.65\linewidth]{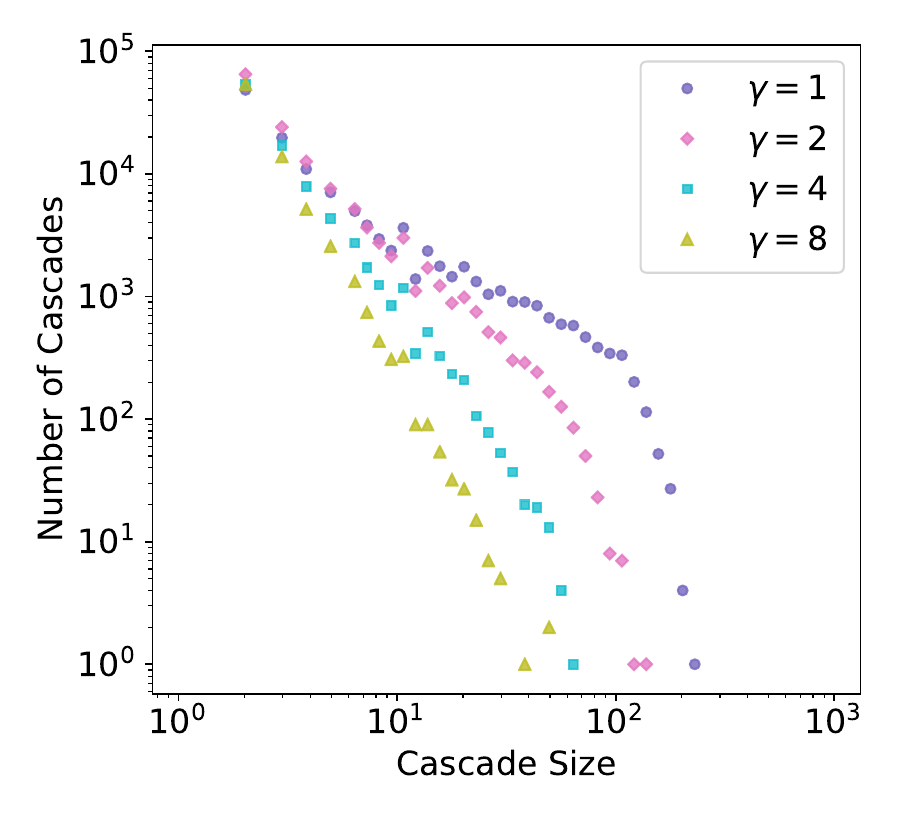}  
    \caption{Cascade size distributions of synthetic cascades generated under different $\gamma$ settings.}
    \label{fig:rt-syn}
\end{figure}

We then generate the spreading trace of each cascade $c$ by simulating an SIR process under the following setting: initial infection rate $\rho=1/N$, cascade-specific edge-wise transmission rates $\lambda_{ij}^c=\sum_{k=1}^K \pi_k^c \alpha_{ij}^k$, ending time $T=10$, and recovery rate $\gamma$ taking 1, 2, 4, or 8. By varying the recovery rate $\gamma$ as such, we are able to cover a relatively wide range of spreading settings with different cascade size distributions. Intuitively, the larger the $\gamma$, the more difficult the cascades will spread to a broader set of nodes. Figure~\ref{fig:rt-syn} shows the cascade size distributions varied by $\gamma$ under the core network setting. We can see that the case of $\gamma=2$ best matches the \textit{ClimateSkepticCascades} dataset, and the case of $\gamma=8$ best matches the \textit{PoliSciCascades} dataset.

After simulating all cascades, we remove the uninformative ones where only one node is activated and no spreading is observed.

\subsection{Technical Settings}
The implementation of our inference program consists of a single layer phase and a multilayer phase. In our testing, after the single layer inference that returns the estimated edge weights of the aggregated network $G_A$, we rank the edges by the estimated edge weights in descending order, and take the top $1.1\cdot|E_A|$ edges as the inferred edges $E_S$, where $|E_A|$ is the number of edges in the ground-truth aggregated network. We expect to restrict memory usage by setting this limit, yet meanwhile allow a decent level of error tolerance.

We measure the accuracy of the single layer inference by the area-under-curve (AUC) score \cite{fawcett2006introduction} of the estimated edge weights on the aggregated network, which assesses the overall classification accuracy of edge existence across all possible thresholds of edge weights. For the multilayer inference, we measure respectively the classification accuracy of the cascade layers (abbreviated as $\pi$ accuracy), and the Spearman's rank correlation \cite{spearman1904proof} between the inferred and ground-truth layer-wise edge transmission rates of the non-zero entries of $\boldsymbol{\alpha}$ (abbreviated as $\alpha$ correlation).

We use for single layer inference the Adam optimizer \cite{kingma2014adam} with an initial learning rate of 0.5, minimum 100 iterations, and maximum 500 iterations; for multilayer inference, we use the Adam optimizer with an initial learning rate of 0.1, minimum 100 iterations, and maximum 3000 iterations. Additionally, we monitor the percentage decrease of the objective function value at each iteration, and stop the single layer inference when the value decreases less than 0.01\%, and the multilayer inference when the value decreases less than 0.0001\%. We run each multilayer inference 3 times with respectively the randomization seed of 0, 1, and 2, and take the best set of results across the three runs as measured by the $\pi$ accuracy. All tests are run on an NVIDIA Tesla P100 GPU card with 3854 threads and 16GB memory.

\subsection{Validation Test}
\label{sec:comparison}
Before conducting the experiments on our implementation of the inference framework, we validate its effectiveness by evaluating the performance of our code against existing implementations of multilayer diffusion network inference, including \textsc{MixCascades}, \textsc{MMRate}, and \textsc{FASTEN}. We compare the inference accuracy and runtime of the implementations when respectively applied to three synthetic datasets generated under our model, and three other generated under the \textsc{FASTEN} model. Specifically, we use our data generated under the setting of $N=1000, \phi=0, \mu_{in}=0.5, \epsilon_{max}=0, \gamma=2$, $K$ taking 2, 3, or 4; and we use $16\cdot|E_A|$ cascades for the inference (where $|E_A|$ is the number of edges in the aggregated single layer network), while in the multilayer phase we filter out cascades of size below or equal to 8 (more about the motivation of cascade filtering in \textit{Varying Cascade Filtering}). Additionally, we use \textsc{FASTEN} data generated under the exponential transmission time model on a network of 1024 nodes, 3 layers, 2048 edges per layer, and respectively a random, hierarchical, or core-periphery network structure (with a parameter matrix of respectively $[0.5, 0.5; 0.5, 0.5]$, $[0.9, 0.1; 0.1, 0.9]$, or $[0.9, 0.5; 0.5, 0.3]$ for the Kronecker graph generator). The edge transmission rates are sampled uniformly from $(0,1)$ for \textsc{FASTEN} data. 

We run our implementation on an NVIDIA Tesla P100 GPU card with 3854 threads and 16GB memory, and run the other implementations using 10 cores from an Intel Xeon Gold 6248 @ 2.50 GHz Processor, with a hyperparameter setting as reported in the \textsc{FASTEN} paper. For the validation test in specific, we stop the optimization process of \textsc{MultiC} at respectively 25, 50, and 100 iterations, to observe the growth of its inference accuracy with its runtime. Additionally, we evaluate the inference accuracy of the implementations by the average precision-recall AUC (abbreviated as PR AUC) of inferred edge transmission rates on each layer, which is the metric proposed in the \textsc{FASTEN} paper that measures the accuracy of the inferred multilayer network.

\begin{figure}[t]
    \centering
    \includegraphics[width=\linewidth]{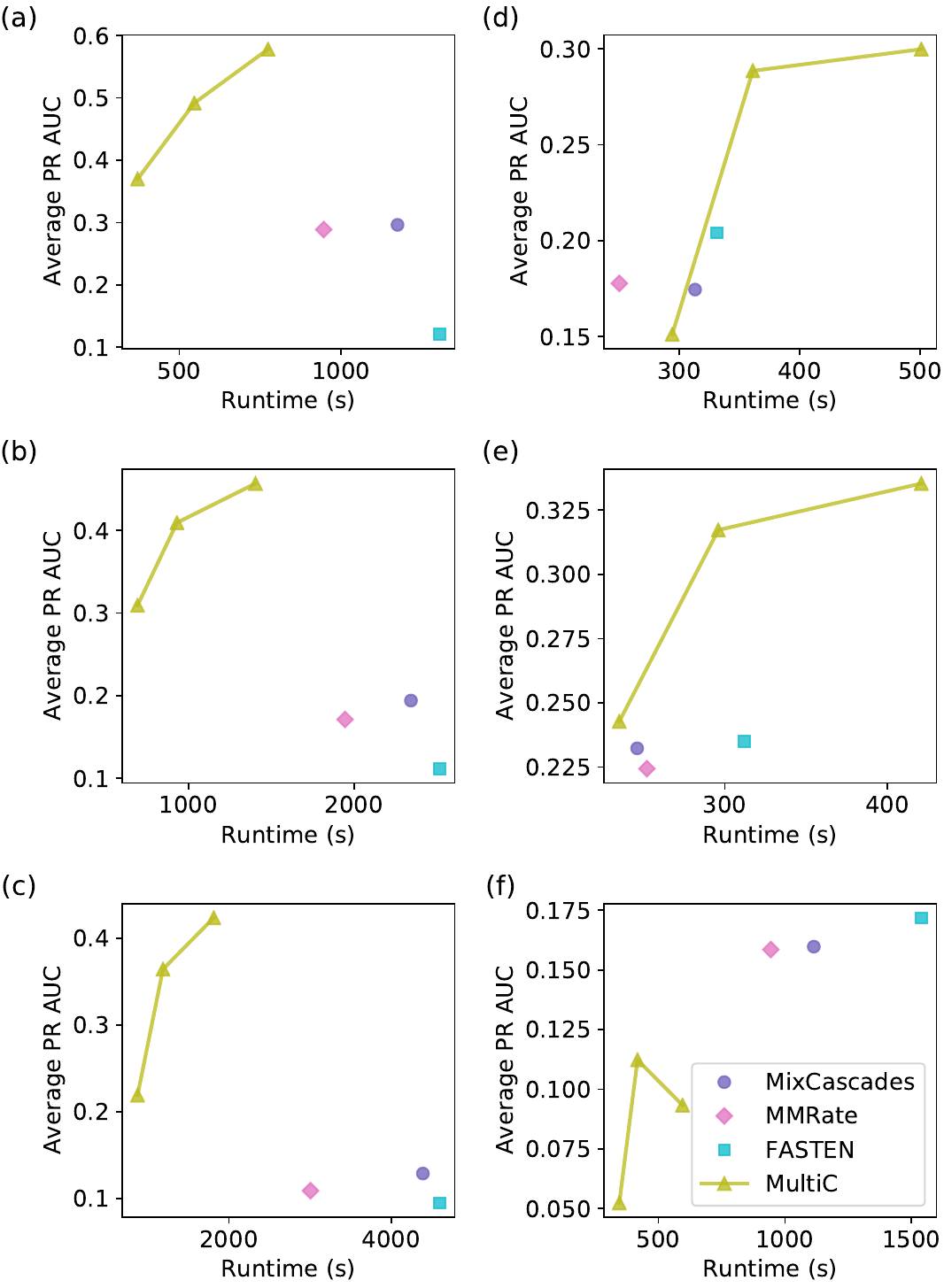}  
    \caption{Inference accuracy of \textsc{MultiC} compared with \textsc{MixCascades}, \textsc{MMRate}, and \textsc{FASTEN}, on synthetic data generated respectively under the setting of (a) \textsc{MultiC}, $K=2$, (b) \textsc{MultiC}, $K=3$, (c) \textsc{MultiC}, $K=4$, (d) \textsc{FASTEN}, random, (e) \textsc{FASTEN}, hierarchical, and (f) \textsc{FASTEN}, core-periphery.}
    \label{fig:comp}
\end{figure}

As plotted in Figure~\ref{fig:comp}, the results show that on all of our datasets, our implementation achieves significantly higher accuracy than previous ones in shorter runtime. Remarkably, on two of the \textsc{FASTEN} datasets where the data generation model is different from what we assume, our implementation also achieves higher accuracy within comparable runtime. 

The superiority of our implementation in inference accuracy is probably due to the fact that the other implementations use stochastic gradient descent for optimization, so that in each of their iterations only a part of the data is used for calculating the gradient; by contrast, all data is used to inform the direction of optimization in every of our iterations, so that our implementation should converge better to the optimum. It usually takes very long time to run the optimization with full data used in every iteration, but we have managed to achieve with GPU computing a comparable runtime as previous implementations.

\section{Results}
\label{sec:results}
Using the inference framework and synthetic data described above, we systematically assessed the performance of the inference framework under different network and diffusion settings. Specifically, we looked at how the inference accuracy varies with cascade size distribution, cascade filtering, network density, network size, number of layers, layer overlap, and mixed spreading. We report these results in the following paragraphs.


\subsection{Varying Cascade Size Distribution} In our first experiment, we have the setting of $N=1000, \mu_{in}=0.5, K=2, \phi=0, |E_A|=4422, \epsilon_{max}=0$, no cascade filtering, and $\gamma$ taking 1, 2, 4, or 8. Under each $\gamma$ value, we inspect how the inference performance changes as we increase the number of cascades, more specifically when the cascade-edge ratio (i.e., the ratio between the number of cascades $C$ and the number of edges in the aggregated network $|E_A|$, abbreviated as C-E ratio) is 1, 2, 4, 8, or 16. 

\begin{figure}
    \centering
    \includegraphics[width=\linewidth]{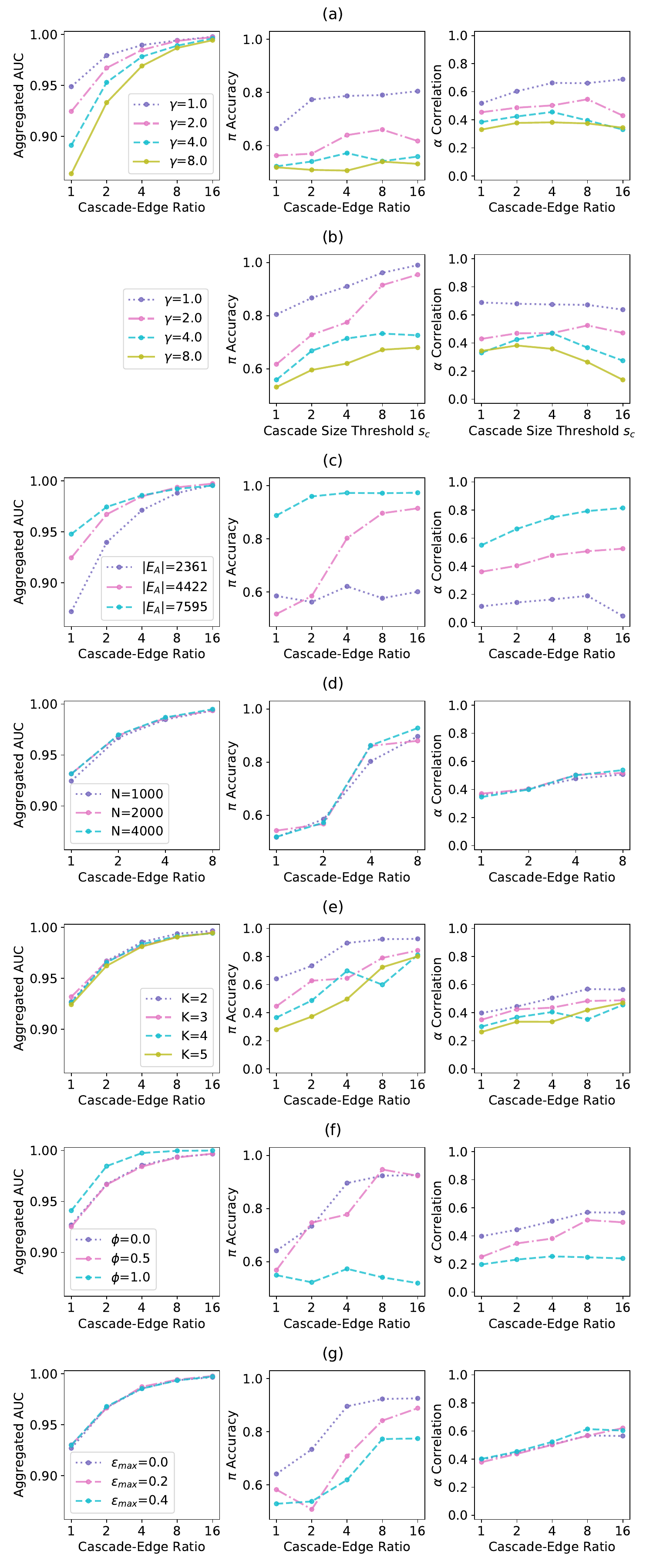}  
    \caption{Experimental results of inference accuracy varied with respectively (a) cascade size distribution, (b) cascade filtering, (c) network density, (d) network size, (e) number of layers, (f) layer overlap, and (g) level of mixed spreading.}
    \label{fig:res}
\end{figure}

The results are shown in Figure~\ref{fig:res}a. We see that the accuracy of the single layer inference almost always grows with the number of cascades used, but it remains lower for larger $\gamma$ values. This is reasonable because cascades generated under larger $\gamma$ values are of smaller sizes, and thus contain less information for the edge inference. However, regardless of the $\gamma$ value, the AUC score always exceeds 0.99 when provided $16\cdot|E_A|$ cascades. We provide in Table~\ref{tab:auc} a mapping from the AUC score to the recovery rate of edges in the ground-truth network, which serves as a reference for interpreting the values. For example, under the setting of $N=1000, |E_A|=4422, |E_S|=1.1\cdot|E_A|=4864, \gamma=8$ and C-E ratio=4, the AUC score is approximately 0.97, and the single layer inference is able to discover about 89\% of the ground-truth edges, among the 4864 edges it infers.

\begin{table}[t]
    \centering
    \begin{tabular}{p{2cm} R{1.1cm} R{3cm}}
    \toprule
     & AUC & Edge Recovery Rate \\
    \midrule
    C-E Ratio=1 & 0.863 & $2989/4422=67.6\%$\\
    C-E Ratio=2 & 0.933 & $3567/4422=80.7\%$ \\
    C-E Ratio=4 & 0.969 & $3938/4422=89.1\%$ \\
    C-E Ratio=8 & 0.987 & $4113/4422=93.0\%$ \\
    C-E Ratio=16 & 0.994 & $4257/4422=96.3\%$ \\
    \bottomrule
    \end{tabular}
    \caption{Mapping from the AUC score to the recovery rate of ground-truth edges ($N=1000,|E_A|=4422,|E_S|=4864,\gamma=8$).}
    \label{tab:auc}
\end{table}

With respect to the multilayer inference accuracy, we observe that when $\gamma=4$ or $\gamma=8$ and there is no cascade filtering, the $\pi$ accuracy and $\alpha$ correlation are not significantly better than baseline (0.5 and 0 respectively), and both metrics do not necessarily increase with the number of cascades used. At best, when $\gamma=1$, the $\pi$ accuracy reaches about 80.5\%, and the $\alpha$ correlation reaches 0.566, given $16\cdot|E_A|$ cascades. In a more realistic case of $\gamma=2$, the best $\alpha$ correlation is 0.373 and the best $\pi$ accuracy is only 66.0\%.

\subsection{Varying Cascade Filtering}
\label{res:filtering}
In our second experiment, we explore if excluding small cascades helps improve the multilayer inference accuracy. The idea stems from the observation that an extra cascade layer membership variable needs to be inferred in the multilayer inference when a new cascade is added to the data; for the entire inference system, it is not clear whether this extra burden brought by a small cascade will outweigh the extra information it contributes. We test specifically under the setting of $N=1000, \mu_{in}=0.5, K=2, \phi=0, |E_A|=4422, \epsilon_{max}=0, \text{C-E ratio}=16$, $\gamma$ taking 1, 2, 4, or 8, and cascade size threshold $s_c$ taking respectively 1, 2, 4, 8, or 16. With a certain threshold $s_c$, all cascades of size below or equal to $s_c$ will be excluded in the inference. Note that taking the threshold of $s_c=1$ is equivalent to no filtering because we have already removed cascades of size 1 (i.e., those with no spreading).

We find that filtering out small cascades effectively improves the accuracy of the multilayer inference, as shown in Figure~\ref{fig:res}b. Overall, we can observe significant improvement in $\pi$ accuracy under all $\gamma$ settings, but especially under $\gamma=1$ and $\gamma=2$. Specifically, as the cascade size threshold grows from 1 to 16, $\pi$ accuracy grows from 80.5\% to 99.0\% under $\gamma=1$, and from 61.7\% to 95.5\% under $\gamma=2$. On the other hand, $\alpha$ correlation mostly fluctuates under all $\gamma$ settings, which potentially reflects the fluctuating level of balance in the entire inference system as cascade-wise information increases and the total number of cascades decreases. Although no monotonic trend can be observed, the good news is that compared with the baseline where no filtering is performed, there exist cascade size thresholds for all $\gamma$ values under which $\pi$ accuracy is significantly higher than baseline, while $\alpha$ correlation also increases upon baseline or at least stays at the same level. For different downstream tasks, the most appropriate cascade size threshold can be chosen based on a weighted evaluation of both metrics.

\subsection{Varying Network Density}
We conduct our third experiment under the setting of $N=1000, K=2, \phi=0, \epsilon_{max}=0, \gamma=2, s_c=8$, $\mu_{in}$ taking 0, 0.5, or 1 (consequently, $|E_A|$ taking 2361, 4422, or 7595), and C-E ratio taking 1, 2, 4, 8, or 16. We see in Figure~\ref{fig:res}c that the single layer inference seems to have the worst performance on the sparsest network, given the same cascade-edge ratio; yet with $16\cdot|E_A|$ cascades, the AUC score still exceeds 0.98 in all cases. The multilayer inference also performs better on denser networks both evaluated by $\pi$ accuracy and $\alpha$ correlation, which can be explained by the higher proportion of large cascades in denser networks. It is worth noting that on the sparsest network we have here, the multilayer inference accuracy is extremely low and almost does not increase at all with the number of cascades.

\subsection{Varying Network Size}
We conduct our fourth experiment under the setting of $K=2, \phi=0, \mu_{in}=0.5, \epsilon_{max}=0, \gamma=2, s_c=8$, $N$ taking 1000, 2000, or 4000 (consequently, $|E_A|$ taking 4422, 8707, or 17768), and C-E ratio taking 1, 2, 4, or 8. Figure~\ref{fig:res}d shows that the accuracy of neither the single layer inference nor the multilayer inference differs significantly with network size.

\subsection{Varying Number of Layers}
We conduct our fifth experiment under the setting of $N=1000, \phi=0, \mu_{in}=0.5, \epsilon_{max}=0, \gamma=2, s_c=8$, $K$ taking 2, 3, 4, or 5 (consequently, $|E_A|$ taking 4422, 6797, 8948 or 11082), and C-E ratio taking 1, 2, 4, 8, or 16. We observe in Figure~\ref{fig:res}e that the accuracy of the single layer inference does not vary significantly with the number of layers in the network. Meanwhile, within a certain level of fluctuation, the multilayer inference achieves slightly less accurate results when the number of layer increases. This matches the intuition that it is relatively more difficult to decompose a network into more layers given the same amount of information.

\subsection{Varying Layer Overlap}
In our sixth experiment, we have the setting of $N=1000, K=2, \mu_{in}=0.5, \epsilon_{max}=0, \gamma=2, s_c=8$, $\phi$ taking 0, 0.5, or 1 (consequently, $|E_A|$ taking 4422, 3040, or 2142), and C-E ratio taking 1, 2, 4, 8, or 16. Figure~\ref{fig:res}f shows that the edge existence in the aggregated network is inferred more accurately on a network with full layer overlap, presumably because in the single layer inference, this situation is equivalent to having double number of cascades for inferring the same network. However, the cascade layer membership is inferred extremely inaccurately (i.e., with below 60\% accuracy) on a network with full layer overlap, due to the limited amount of differentiation between the layers (i.e., only layer-wise edge transmission rates are different).  Between half overlap and no overlap settings, there is no significant difference in the accuracy of the edge existence inference or the cascade layer membership inference. On the other hand, the accuracy of the inferred layer-wise edge transmission rates in general decreases with layer overlap.

\subsection{Varying Level of Mixed Spreading}
In our final experiment, we have the setting of $N=1000, K=2, \phi=0, \mu_{in}=0.5, \gamma=2, s_c=8$, $\epsilon_{max}$ taking 0, 0.2, or 0.4, and C-E ratio taking 1, 2, 4, 8, or 16. We find that the accuracy of both edge existence and edge weight inference does not differ significantly with the level of mixed spreading, but the accuracy of the cascade layer membership inference indeed decreases with layer mixing, as shown in Figure~\ref{fig:res}g.

\subsection{Runtime and Memory Usage}
\label{sec:runtime}
According to our empirical testing, the runtime of our single layer inference scales approximately linearly with the number of edges in the network, and the memory usage scales approximately linearly with $\max(|E_P|, N\cdot C)$, as we expected. Meanwhile, both the runtime and the memory usage of the multilayer inference scale approximately linearly with both the number of the edges in the network and the number of cascades used in the multilayer phase. In our most computationally heavy test case with $\gamma=2, s_c=8$, a network with 4000 nodes and 17768 edges when aggregated, and 142144 (resp. 15405) cascades before (resp. after) filtering, the single layer inference finishes 500 iterations in 108 minutes with 3.1 GB of GPU memory usage, and the multilayer inference finishes 3000 iterations in 23 minutes with 13.3 GB of GPU memory usage. As we can see, the runtime of the inference is relatively acceptable on large networks; but for the inference to run within limited GPU memory, one might have to decrease either the number of nodes, the number of edges, or the number of cascades used in the inference, for example by focusing on a smaller subset of nodes and edges, or filtering out relatively uninformative cascades.

\section{Discussion}
\subsection{Findings \& Implications}
Our results first reveal interesting performance dynamics of the multilayer diffusion network inference framework when applied to realistic spreading data. For example, the accuracy of the edge existence inference on the aggregated network always increases with the number of cascades, regardless of any network structure or diffusion setting. Meanwhile, given the same cascade-edge ratio, the accuracy of the inferred layer-wise edge weights and the layer membership of cascades increases with network density, decreases with the number of layers in the network, and regresses to baseline when there are fewer large cascades in the spreading data for inference. More interestingly, we find that excluding small cascades when conducting the multilayer inference significantly increases the classification accuracy of the cascade layers, especially when there exists a sufficient number of large cascades in the data. This is also a very practical improvement of the inference method since it also decreases runtime and memory usage, and thus potentially mitigates the shortage of time and memory resources.

More importantly, we have shown a wide range of cases where the inference accuracy of the multilayer diffusion network is only slightly better than the random guessing baseline, for example when the number of cascades in the data is very limited, or the proportion of large cascades is very low. Our work thereby shows the potential inapplicability of the inference framework to social media data, and highlights the need for carefully evaluating the framework's applicability before trying to infer a multilayer diffusion network from real-world spreading data. Otherwise, the interpretations of the results can be extremely misleading when the solution quality is not guaranteed. Our results and our open-source implementation can serve as a useful tool for this applicability evaluation. By matching the objective real-world setting to the most similar synthetic setting among those we reported, or running our code under a further personalized setting that best mimics the dataset, one can get a rough estimate of how well the inference performs when applied to the target dataset.

\subsection{Limitations \& Future Work}
Our work is limited in the minimality of the generative diffusion model we assume, and the exponential transmission time model we specify. However, our implementation can be easily extended to support diffusion models with extra elements that stay constant during the inference process, as well as power law or Rayleigh transmission time models. Our implementation is also limited in scalability, due to the relatively high consumption of GPU memory in the multilayer phase of inference. The optimization process can potentially be redesigned to improve memory usage, for example by further decomposing the multilayer phase into separate stages, and using only a fraction of the data to inform the optimization in each stage. We leave these extensions of our implementation for future work.

As much as we tried to cover a wide range of realistic network and diffusion settings in our synthetic data experiments, there definitely remain cases that need to be further explored. For example, it will be interesting to examine the performance of the inference framework when the ground-truth diffusion network has different community structures, as information diffusion is shown to be affected by community structure in online social networks \cite{tsugawa2019empirical}. Besides, it remains to be studied how well the inference works under threshold-based complex contagion models, such as the Watts' threshold model \cite{watts2002simple}.

We are also unfortunately not able to show the results of the inference framework when applied to real social media data, because both real datasets we collected match cases of inference inaccuracy. However, this failure exactly reflects how inapplicable the inference framework can be to real datasets. In future work, we expect to inspect a broader range of real datasets, among which we can hopefully find a few that allow the accurate inference of multilayer diffusion networks and the interpretation of the inferred results.

\appendix

\section{Appendix}
\subsection{Derivation of Formulas}
\label{app:drv}
The likelihood of cascade $c$ is
\footnotesize
$$\begin{aligned}
L(&c)=\prod_{j:t_j^c<T} \Gamma_j^+(c) \times \prod_{n:t_n^c>T} \Gamma_n^-(c) \\
=&\prod_{j:t_j^c<T} \left[\sum_{i:t_i^c<t_j^c} f(\Delta t_{ij}^c;\lambda_{ij}^c) \prod_{u:u\neq i,t_u^c<t_j^c} S(\Delta t_{uj}^c;\lambda_{uj}^c)\right] \\
&\times \prod_{n:t_n^c>T} \prod_{m:t_m^c<T} S(T-t_m^c;\lambda_{mn}^c) \\
=&\prod_{j:t_j^c<T} \left[ \left( \sum_{i:t_i^c<t_j^c} f(\Delta t_{ij}^c;\lambda_{ij}^c)  \prod_{u:u\neq i,t_u^c<t_j^c} S(\Delta t_{uj}^c;\lambda_{uj}^c)\right) \right. \\
 &\times \left. \prod_{n: t_n^c>T} S(T-t_j^c;\lambda_{jn}^c)\right] \\
=&\prod_{j:t_j^c<T} \left[\prod_{u:t_u^c<t_j^c} S(\Delta t_{uj}^c;\lambda_{uj}^c)\times \sum_{i:t_i^c<t_j^c} H(\Delta t_{ij}^c;\lambda_{ij}^c) \right. \\
&\times \left. \prod_{n: t_n^c>T} S(T-t_j^c;\lambda_{jn}^c)\right].
\end{aligned}$$
\normalsize

Under the exponential transmission time model, the negative log likelihood of cascade $c$ can be written as
\footnotesize
$$\begin{aligned}
-\log& L(c)=-\sum_{j:t_j^c<T}\left(\sum_{u:t_u^c<t_j^c}\log S(\Delta t_{uj}^c;\lambda_{uj}^c) \right.\\
&\left.+\log\sum_{i:t_i^c<t_j^c} H(\Delta t_{ij}^c;\lambda_{ij}^c)+\sum_{n: t_n^c>T}\log S(T-t_j^c;\lambda_{jn}^c)\right) \\
=&\sum_{j:t_j^c<T}\left(\sum_{u:t_u^c<t_j^c}\lambda_{uj}^c \Delta t_{uj}^c - \log\sum_{i:t_i^c<t_j^c} \lambda_{ij}^c \right.\\
&\left.+ \sum_{n: t_n^c>T}\lambda_{jn}^c (T-t_j^c)\right) \\
=&\sum_{j:t_j^c<T}\left(\sum_{u:t_u^c<t_j^c} \sum_{k=1}^K \Delta t_{uj}^c\pi_k^c \alpha_{uj}^k - \log\sum_{i:t_i^c<t_j^c} \sum_{k=1}^K \pi_k^c \alpha_{ij}^k \right.\\
&\left.+\sum_{n: t_n^c>T}\sum_{k=1}^K (T-t_j^c) \pi_k^c \alpha_{jn}^k\right).
\end{aligned}$$
\normalsize
and the total negative log likelihood of all cascades as
\footnotesize
$$\begin{aligned}
\mathcal{L}(&\mathbf{c};\boldsymbol{\alpha}, \boldsymbol{\pi})=\sum_c \sum_{j:t_j^c<T}\left(\sum_{u:t_u^c<t_j^c} \sum_{k=1}^K \Delta t_{uj}^c\pi_k^c \alpha_{uj}^k \right.\\
&\left.- \log\sum_{i:t_i^c<t_j^c} \sum_{k=1}^K \pi_k^c \alpha_{ij}^k+\sum_{n: t_n^c>T}\sum_{k=1}^K (T-t_j^c) \pi_k^c \alpha_{jn}^k\right).
\end{aligned}$$
\normalsize

\subsection{Formulation of Inference Phases}
\label{app:phases}
The single layer phase of inference in our implementation can be formulated as solving the optimization problem
\footnotesize
$$\begin{aligned}
\underset{\boldsymbol{\alpha'}}{\text{minimize}}\quad &\sum_c \sum_{j:t_j^c<T}\left(\sum_{\substack{u:t_u^c<t_j^c\\ (u,j)\in E_P}} \Delta t_{uj}^c\alpha_{uj}' - \log\sum_{\substack{i:t_i^c<t_j^c\\ (i,j)\in E_P}} \alpha_{ij}' \right. \\
&+ \left. \sum_{n: t_n^c>T} (T-t_j^c) \alpha_{jn}' \right) \\
\text{subject to\quad} & 0\leq \alpha_{ij}'\leq 1 \text{ for all } i \text{, for all } j,
\end{aligned}$$
\normalsize
where $\alpha'_{ij}$ is the edge indicator variable in the aggregated network. We derive the set of edges that exist in the aggregated network, $E_S$, by setting a threshold $\delta$, and assuming $(i,j)$ exists in $E_S$ if and only if $\alpha'_{ij}>\delta$. 

Meanwhile, the multilayer phase solves the optimization problem
\footnotesize
$$\begin{aligned}
\underset{\boldsymbol{\alpha}, \boldsymbol{\pi}}{\text{minimize}}\quad &\sum_c \sum_{j:t_j^c<T}\left( \sum_{\substack{u:t_u^c<t_j^c\\ (u,j)\in E_S}} \sum_{k=1}^K \Delta t_{uj}^c\pi_k^c \alpha_{uj}^k \right. \\
&- \log\sum_{\substack{i:t_i^c<t_j^c \\ (i,j)\in E_S}} \sum_{k=1}^K \pi_k^c \alpha_{ij}^k\\
&+\left. \sum_{n: t_n^c>T}\sum_{k=1}^K (T-t_j^c) \pi_k^c \alpha_{jn}^k \right) \\
\text{subject to\quad} & 0\leq \pi_k^c\leq 1 \text{ for all } k \text{, for all } c \\
& \sum_1^K \pi_k^c=1 \text{ for all }c \\
& 0\leq \alpha_{ij}^k\leq 1 \text{ for all } i \text{, for all } j \text{, for all } k.
\end{aligned}$$
\normalsize

\section*{Ethical Statement}
From a broader perspective, our work further connects the technical literature of multilayer diffusion network inference to their empirical applications on real social media data. Such empirical studies may help uncover nuanced homogeneity and heterogeneity in the spreading of online content, and thus advance our understanding of information propagation and opinion dynamics in online social networks. In particular, they could reveal a richer spreading context of misleading or harmful content, and inspire designs of more effective regulation measures.

On the other hand, by showcasing the potential use of the inference framework on social media data and providing an open-source implementation of it, our work potentially induces a risk for people to apply the framework on unanonymized social media data, consequently inferring the connection between real users and how ideas spread among them. We therefore encourage the users of the inference framework to always apply it to spreading data with identifying information removed, and interpret the revealed spreading dynamics from a content-oriented perspective.

\section*{Acknowledgments}
We would like to thank the Aalto Science-IT project for their computational resources, the Aalto SciComp team for their extremely helpful technical support, and Bisbee, Larson, and Munger \cite{bisbee2020polisci} for sharing their dataset with us. Our appreciation also goes to  Barbara Keller for her insightful comments on the manuscript. All the authors acknowledge funding from the Academy of Finland, project ECANET, number 320781. In addition, TC acknowledges Academy of Finland, project ECANET, number 320780.

\bibliography{references}

\end{document}